\documentclass[twocolumn,showpacs,showkeys,prc,aps,amsmath,amssymb,10pt]{revtex4-1}
\usepackage{graphicx}
\usepackage{color}
\usepackage{multirow}
\usepackage{longtable}
\begin{document}

\title{Shell model studies of the $^{130}$Te neutrinoless double-beta decay} %Title of paper

% repeat the \author .. \affiliation  etc. as needed
% \email, \thanks, \homepage, \altaffiliation all apply to the current author.
% Explanatory text should go in the []'s, 
% actual e-mail address or url should go in the {}'s for \email and \homepage.
% Please use the appropriate macro for the type of information

% \affiliation command applies to all authors since the last \affiliation command. 
% The \affiliation command should follow the other information.

\author{Andrei Neacsu}
\email{neacs1a@cmich.edu}
\author{Mihai Horoi}
\email{mihai.horoi@cmich.edu}
\affiliation{Department of Physics, Central Michigan University, Mount Pleasant, Michigan 48859, USA}

\date{\today}

\begin{abstract}
Most uncertainties regarding the theoretical study of the neutrinoless double-beta decay are related to the accuracy of the nuclear matrix elements that appear in the expressions of the lifetimes. 
We calculate the nuclear matrix elements for the $0 \nu \beta \beta$ decay 
of $^{130}$Te in a shell model approach, using a recently proposed effective Hamiltonian. 
 To ensure the reliability of the results, we investigate this Hamiltonian by performing calculations of  spectroscopic quantities and comparing them to 
the latest experimental data available, and we analyze the $2 \nu \beta \beta$ and the $0 \nu \beta \beta$ decay nuclear matrix elements of $^{136}$Xe. Finally, we report new nuclear matrix  
for the $^{130}$Te considering the 
light neutrino exchange and heavy neutrino exchange mechanisms, alongside with an overview of some recent values reported in the literature. 
\end{abstract}

%14.60.Pq	Neutrino mass and mixing (see also 12.15.Ff Quark and lepton masses and mixing)
%21.60.Cs	Shell model
%23.40.-s	β decay; double β decay; electron and muon capture
%23.40.Bw	Weak-interaction and lepton (including neutrino) aspects (see also 14.60.Pq Neutrino mass and mixing)
%\keywords{Neutrino mass, Shell model, Double beta decay, Nuclear matrix elements}
\maketitle 

\section{Introduction}
Neutrinoless double-beta decay ($0 \nu \beta \beta$) represents a promising beyond Standard Model (BSM) process for studying the lepton number violation (LNV) effects at low energies 
and for understanding neutrino properties, especially the neutrino mass scale and whether neutrino is a 
Dirac or a Majorana fermion \cite{sv82}. 
Neutrino oscillation experiments have successfully measured the squared mass differences among neutrino flavors \cite{KAE10}-\cite{KML11}, yet the nature of the neutrinos
and the absolute neutrino mass cannot be obtained from such measurements. This has led to both theoretical and experimental efforts dedicated to the discovery of the $0 \nu \beta \beta$ decay mode, as reflected in recent 
reviews on the subject \cite{AEE08}-\cite{VOG12}. The most studied mechanism is the exchange of light Majorana neutrinos in the presence of left handed (LH) weak interaction, but other possible mechanisms 
 contributing to the total $0 \nu \beta \beta$ decay rate are taken into consideration. Such 
 mechanisms include the exchange of right-handed heavy neutrinos 
\cite{MOH75},\cite{DOI83}, and mechanisms involving SUSY particles \cite{VES12},\cite{HIR96}. 
To date, the $0 \nu \beta \beta$ decay and the analysis of the same-sign dilepton decay 
channels at hadron colliders \cite{LHC-LNV} are the best approaches to investigate these matters. Complementary information regarding the neutrino physics parameters can be obtained from large-baseline and new reactor neutrino 
oscillation experiments \cite{snowmass}, and from cosmology \cite{cosmo}. 

Recent interest in $^{130}$Te for $0\nu\beta\beta$ decay experiments, such as CUORE \cite{cuore}, presents a pressing need for very accurate evaluation of the $0 \nu \beta \beta$ nuclear matrix elements (NME)
 for this nucleus. Accurate NME are essential for guiding the experimental effort, for comparing with the experimental results of other decaying isotopes, and ultimately for extracting information about the decay mechanism, the neutrino mass scale, and the Majorana CP-violation phases.

The $0\nu\beta\beta$ lifetimes are usually expressed as a product of a leptonic phase space factor (PSF), a NME that depends on the nuclear structure of 
the parent and that of the daughter nuclei, and a LNV parameter related to the BSM mechanism considered. Precise calculations of the PSF and NME, 
together with accurate measurements of the $0\nu\beta\beta$ decay lifetimes, are all needed in order to obtain reliable limits for the LNV parameter. 

The largest discrepancies in the theoretical studies of $0\nu\beta\beta$ decays are related to the calculated values of the NME that are currently investigated by several methods, 
of which the most employed ones are proton-neutron Quasi Random Phase Approximation (pnQRPA) \cite{tu99}-\cite{qrpa-tu-2014}, Interacting Shell Model (ISM) \cite{Cau95}-\cite{ns-jpg41}, 
Interacting Boson Model (IBM-2) \cite{BI09}-\cite{iba-prc2013}, Projected Hartree Fock Bogoliubov (PHFB) \cite{RAH13}, Energy Density Functional (EDF) \cite{RMP10}, and the Relativistic 
Energy Density Functional (REDF) \cite{ring-bmf-2014} method. 
There are still large differences among the NME calculated with different methods and  by different groups, which has been a topic of many debates in the literature 
(\cite{FAE12}-\cite{VOG12}). Recent calculations of the PSF factors \cite{iachello-psfs1},\cite{stoica-mirea-prc88} have been performed with higher accuracy, and differences were found when compared to the older calculations 
%\cite{PR59},
\cite{SC98}.

An important ingredient for accurate shell model calculations of nuclear structure and decay properties of nuclei is the nucleon-nucleon interaction.
%, which dictates the behavior of the mean field as a function of nucleons in the valence shells\cite{zuker_prc_1996},\cite{zuker_prc_1999}.
Realistic effective nucleon-nucleon ($nn$) interactions, derived from free $nn$ potentials, form the microscopic basis of shell model calculations \cite{jensen-kuo}. However, these effective interactions often require additional fine-tuning to the 
available data to gain real predictive power.
In this paper, we  use a recently proposed shell model effective Hamiltonian (called SVD here) \cite{svd_int}
for nucleons between the $Z,N \in \left[50,82 \right] $ shell closures with $0g_{7/2}, 1d_{5/2}, 1d_{3/2}, 2s_{1/2}$ and $0h_{11/2}$ orbitals (called the $jj55$ model space).
Before turning to the calculation of the $^{130}$Te $0\nu\beta\beta$ decay NME, we perform a series of nuclear structure tests of this Hamiltonian for the nuclei involved in the double-beta decays of $^{130}$Te and the nearby
$^{136}$Xe. These tests include the energy spectra, the $B(E2)\uparrow$ transition probabilities, the $GT$ strengths, and the occupation probabilities for both neutrons and protons. In a second step, we reanalyze the double-beta decay NME
for $^{136}$Xe, which was recently described in the large $jj77$ model space that includes all spin-orbit partner orbitals. Unfortunately, the analysis of $^{130}$Te double-beta decay NME in the larger $jj77$ model space is not yet feasible, and we
try to asses if the restriction to the $jj55$ model space, and the use of the SVD fine-tuned shell-model Hamiltonian, can provide reasonable results.
After evaluating the reliability of this Hamiltonian in the reduced $jj55$ model space, 
we calculate the $^{130}$Te NME for both the light neutrino and the heavy neutrino exchange mechanisms. 
We then compare our shell-model results to the most recent NME results from other groups.

The paper is organized as follows. In the following Section we briefly present the formalism for NME involved in the expressions of $0 \nu \beta \beta$ decay lifetimes via exchange of both 
light Majorana neutrinos and heavy neutrinos mechanisms. Section \ref{detail} presents a detailed study of the SVD 
nucleon-nucleon effective interaction for use in shell model calculations for $^{136}$Xe and $^{130}$Te. Subsection \ref{spectra} displays the theoretical spectra, Subsection \ref{be2} 
the calculated $B(E2)\uparrow$ values and Subsection \ref{gt} the evaluated GT strengths. In subsection \ref{nme-xe} we show the NME obtained using the 
SVD Hamiltonian for $^{136}$Xe and an analysis of the contributing components. The $^{130}$Te NME are shown in Section \ref{nme-te}, followed by discussion on the NME from
different groups in Section \ref{discussions}. Section \ref{conclusions} is dedicated to conclusions regarding the use of this new interaction, as well as remarks on the new NME obtained. 

\section{$0 \nu \beta \beta$ decay NME formalism}

Considering the exchange of light left-handed neutrinos and heavy right-handed neutrinos, the following expression for $0 \nu \beta \beta$ 
decay half-lives is a good approximation \cite{MH13}:
\begin{equation}
\left[ T^{0\nu}_{1/2} \right]^{-1}=G^{0\nu} \left( \left| M^{0\nu}_\nu\right|^2 \left| \eta_{\nu L} \right|^2 + \left| M^{0\nu}_N\right|^2 \left| \eta_{N R} \right|^2\right) .
\end{equation}
Here $G^{0\nu}$ is the phase space factor for this decay mode \cite{iachello-psfs1}-\cite{SC98} that depends on the energy decay and nuclear charge (our $G^{0\nu}$ includes the $g^4_A$ factor), $M^{0\nu}_{\nu,N}$ are the the NME, $\eta_{\nu L}$ and $\eta_{N R}$ are the 
neutrino physics parameters associated to the light neutrino exchange and the heavy neutrino exchange mechanisms, respectively.
The neutrino physics parameters are \cite{FAE11,MH13}:
\begin{equation}
 \eta_{\nu L} =\! \sum_k^{light} U_{ek}^2 \frac{m_k }{m_e}, 
\ \eta_{N R}=\! \left(\frac{M_{W_L}}{M_{W_R}}\right)^4\sum_k^{heavy} V_{ek}^2 \frac{m_p }{M_k},
\end{equation}
with $m_e$ being the electron mass and $m_p$ the proton mass. Here we assume that the neutrino mass eigenstates are separated as light, $m_k(m_k \ll 1 \ eV) $,  and heavy, $M_k (M_k \gg 1 \ GeV)$.
$U_{ek}$ and $V_{ek}$ are electron neutrino mixing matrices for the light left-handed and heavy right-handed neutrino, respectively \cite{MH13},\cite{rod_jhep_1309}.
Using the experimental lifetimes data from two $0 \nu \beta \beta$ decaying nuclei and assuming only contributions from these two non-interfering decay mechanisms, it is possible to obtain 
information about the neutrino physics parameters $\eta_{\nu L}$ and $\eta_{N R}$ \cite{FAE11}.  

The expressions for $M^{0\nu}_{\nu,N}$ have the following structure:
\begin{equation}
 M^{0\nu}_{\nu,N}=M^{0 \nu}_{GT}-\left( \frac{g_V}{g_A} \right)^2 \cdot M^{0 \nu}_F + M^{0 \nu}_T \ ,
\end{equation}
where $g_V$ and $g_A$ are the Vector and the Axial-Vector coupling strengths, respectively, while $M^{0 \nu}_{GT}$, $M^{0 \nu}_F$ and  $M^{0 \nu}_T$ are the Gamow-Teller ($GT$), the Fermi($F$) and the Tensor($T$) components, respectively, defined as follows:
\begin{equation}
M_\alpha^{0\nu} = \sum_{m,n} \left< 0^+_f\| \tau_{-m} \tau_{-n}O^\alpha_{mn}\|0^+_i \right> \ ,
\end{equation}
where $O^\alpha_{mn}$ are transition operators ($\alpha=GT,F,T$) and the summation is over all the nucleon states. 
Explicit expressions for $M_\alpha^{0\nu}$ can be found in several papers, for example Ref. \cite{senkov-horoi-brown-prc2014}.
There are two conventions when dealing with the sign of the tensor contribution \cite{SIM08}, which arise from the sign of 
the second order Bessel function $j_{2}(qr)$ that appears in the radial part of the transition operator. Here we use the convention with the negative sign of $j_{2}(qr)$.

In our calculations, we have employed all the nuclear structure ingredients and parameters used in the recent literature \cite{sim-09}. Our method includes short range correlations (SRC), finite nucleon size effects (FNS), and higher order corrections of the nucleon current (HOC) \cite{HS10}-\cite{ns-jpg41}.

\section{Analysis and validation of the effective interaction}\label{detail}

For the shell model calculation of the $^{130}$Te $0\nu\beta\beta$ decay NME, we need a suitable effective Hamiltonian, one which can reliably describe the structure of the nuclei involved in the decay. 
One option is to use an effective Hamiltonian in a large model space that includes the $0g_{7/2}, 0g_{9/2}, 1d_{5/2}, 1d_{3/2}, 2s_{1/2}, 0h_{9/2}$ and $0h_{11/2}$ orbitals (called $jj77$). 
This approach was successfully used in the case of $^{136}$Xe \cite{hb_prl_2013}. However, the shell-model dimensions dimension for $^{130}$Te in this model space are too large for realistic shell model calculations.
Another option is using the $jj55$ model and an effective Hamiltonian fine tuned to the experimental data. The effects of decreasing the model space by excluding the spin-orbit partners orbitals $0g_{7/2}$ and $0h_{11/2}$ are presented and discussed in this paper.
We investigate the SVD Hamiltonian reported in \cite{svd_int}, which was fine tuned using the experimental data for Sn isotopes, and we analyze how accurately it describes the nuclei of interest in this study by comparing to the available experimental data. This effective
Hamiltonian was obtained starting with a realistic CD-Bonn $nn$ potential \cite{machleidt-prc-2001}, and the core-polarization effects \cite{jensen-pr-1995}
have been taken into account by renormalizing the interaction using the perturbative $G$-matrix approach. Ref. \cite{svd_int} presents a detailed study of this effective Hamiltonian for 
Sn isotopes.

Using this effective Hamiltonian, we calculate and compare with the experimental data, when available, the following spectroscopic quantities for the nuclei in the region of interest: 
the energy spectra for the first $\left[ 0^+ - 6^+ \right]$ states, $B(E2)\uparrow$ transition probabilities, occupation probabilities and the Gamow-Teller strengths. 
Finally, to properly validate the interaction for the calculation of the $^{130}$Te NME, we re-analyze the $2\nu\beta\beta$ and the $0\nu\beta\beta$ decay NME for $^{136}$Xe. 

\subsection{Spectra} \label{spectra}

The first study we perform with the SVD Hamiltonian is the comparison of the calculated 
energy spectra of 
$^{130}$Te, $^{130}$Xe, $^{136}$Xe and $^{136}$Ba with the experimental data available. Some of the energy levels could not be accurately identified in the experimental data, thus we omit them from our comparison. 

 \begin{figure} [h!]
 \includegraphics[width=0.95\linewidth]{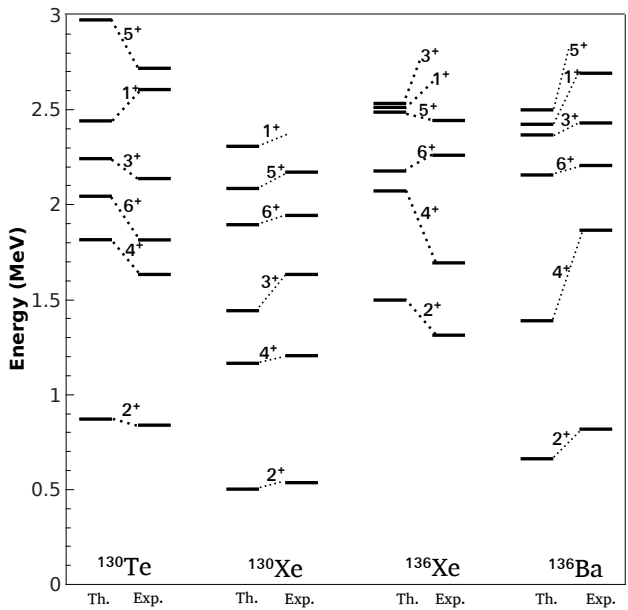}
 \caption{The calculated energy levels for $^{130}$Te, $^{130}$Xe,  $^{136}$Xe and $^{136}$Ba (left columns) compared to the experimentally determined ones (right columns).}
 \label{elevels}
 \end{figure}

Fig. \ref{elevels} presents the low-energy spectra calculated with the SVD Hamiltonian compared the experimental data. In the case of the initial nuclei,$^{130}$Te and $^{136}$Xe, our shell model calculations 
tend to overestimate the experimental levels, while for the final nuclei, $^{130}$Xe and $^{136}$Ba, our results are often below the experimental ones, but within few hundred keV.
We find this effective Hamiltonian to provide satisfactory results concerning the prediction of energies of the initial and the final nuclei.

\subsection{$B(E2)\uparrow$ transitions} \label{be2}

For the $B(E2)\uparrow$ values calculated using the SVD Hamiltonian we use the canonical neutron and proton charges ($e^{eff}_n=0.5$, $e^{eff}_p=1.5$), and we compare with the 
 adopted data \cite{ad-be2}. The study of the the $B(E2)\uparrow$ values done in Ref. \cite{svd_int} uses different neutron effective 
charges ($e^{eff}_n=0.88e$), but it is only for tin isotopes with no protons in the valence shell.
The comparison between our calculated values and the adopted ones can be seen in Table \ref{tab_be2}, where we show the $B(E2)\uparrow$ values for seven nuclei of interest in our study. We notice the very good 
agreement between theory and experiment for all important cases: $^{130}$Te, $^{130}$Xe, $^{136}$Xe and $^{136}$Ba.

\begin{table} [h!]
 \caption{The calculated $B(E2)\uparrow$ values (first row) compared to the adopted ones second row).}
 \begin{tabular}{l|rrrrrrr} \hline \hline
			&$^{128}$Te	&$^{130}$Te	&$^{132}$Te	&$^{130}$Xe 	&$^{132}$Xe	&$^{136}$Xe	&$^{136}$Ba	\\ \hline 
$B(E2)\uparrow_{th.}$	&0.202		&0.153		&0.085		&0.502		&0.390		&0.215		&0.479	\\									
$B(E2)\uparrow_{ad.}$&0.380		&0.297		&0.207		&0.634		&0.468		&0.217		&0.413	\\ \hline \hline

%$e^{eff}_n=0.88$	&$^{128}$Te	&$^{130}$Te	&$^{132}$Te	&$^{130}$Xe 	&$^{132}$Xe	&$^{136}$Xe	&$^{136}$Ba	\\ \hline 
%$B(E2)\uparrow_{th.}$	&0.452		&0.332		&0.198		&1.026		&0.753		&0.338		&0.837	\\									
%$B(E2)\uparrow_{exp.}$&0.380		&0.297		&0.207		&0.634		&0.468		&0.217		&0.413	\\ \hline
 \end{tabular}
 \label{tab_be2}
 \end{table}
 
\subsection{Occupation probabilities} \label{occupancies}
In order to verify how suitable the effective SVD Hamiltonian is to reliably describe the nuclear structure aspects of the nuclei involved in our calculations,
we also tested how accurately it describes the neutron vacancies for $^{128}$Te, $^{130}$Te, $^{130}$Xe and $^{132}$Xe. Our results  
are compared to the latest experimental data reported in Ref. \cite{kay-vacancies-2013}. 

\begin{figure} [h!]
 \includegraphics[width=0.95\linewidth]{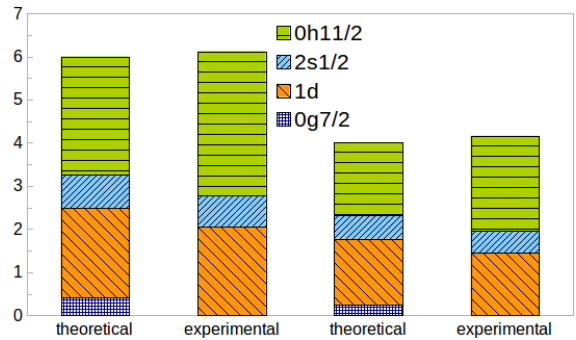}
 \caption{(Color online) Theoretical and experimental \cite{kay-vacancies-2013} neutron shell vacancies for $^{128}$Te and $^{130}$Te.}
 \label{te_neutr_vac}
\end{figure}

Fig. \ref{te_neutr_vac} presents the calculated neutron vacancies in $^{128}$Te and in $^{130}$Te, compared to the experimental results.
One can notice that the predicted neutron vacancy in orbitals $0g_{7/2}$ has not been experimentally confirmed, but also that the experimental sum of the vacancies 
exceeds the exact numbers 6 and 4 in the case of $^{128}$Te and $^{130}$Te, respectively. 
\begin{figure} [h!]
 \includegraphics[width=0.95\linewidth]{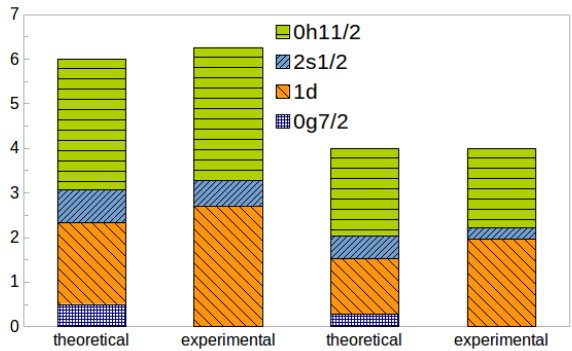}
 \caption{(Color online) Theoretical and experimental \cite{kay-vacancies-2013} neutron shell vacancies for $^{130}$Xe and $^{132}$Xe.}
 \label{xe_neutr_vac}
\end{figure}

Fig. \ref{xe_neutr_vac} shows the comparison between our calculations and the experimental results in the case of $^{130}$Xe and $^{132}$Xe. The same 
observations regarding the $0g_{7/2}$ orbitals are valid for these two nuclei, as for the case of Te isotopes.
We find the agreement among the theoretical and the experimental data satisfactory for the purpose of our calculations.

\begin{figure} [h!]
 \includegraphics[width=0.95\linewidth]{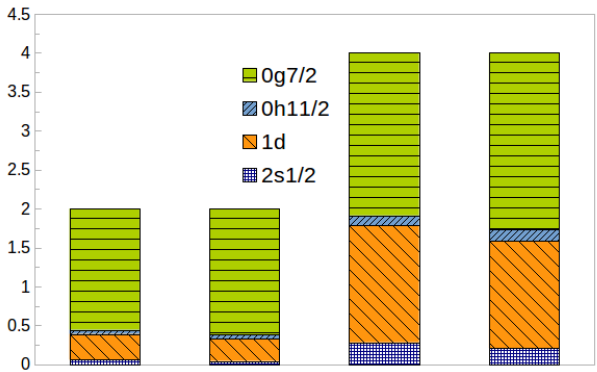}
 \caption{(Color online) Theoretical proton shell occupancies for $^{128}$Te, $^{130}$Te, $^{130}$Xe and $^{132}$Xe.}
 \label{prot_occup}
\end{figure}

In the case of the proton occupancies, there is no reliable experimental data for comparison. We present our theoretical results for proton occupancies in Fig. \ref{prot_occup} for 
$^{128}$Te, $^{130}$Te, $^{130}$Xe and $^{132}$Xe.

\subsection{$GT$ strengths} \label{gt}
\begin{figure} [h!]
 \includegraphics[width=0.95\linewidth]{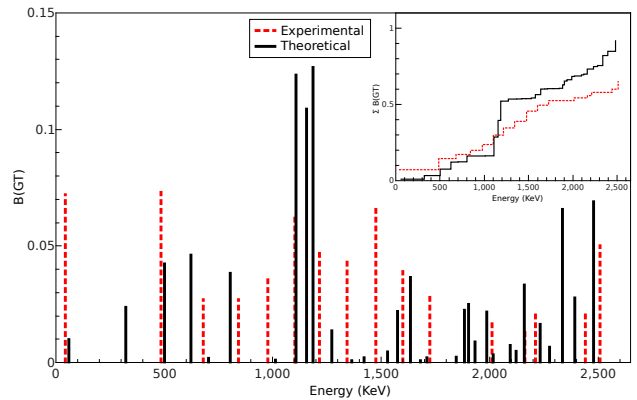}
 \caption{(Color online) Calculated $^{130}$Te GT strengths (solid line) compared to the experimental ones (dotted line)\cite{exp_130te}. The inset presents the calculated and the experimental GT running sum.}
 \label{gt_130te}
 \end{figure}

\begin{figure} [h!]
 \includegraphics[width=0.95\linewidth]{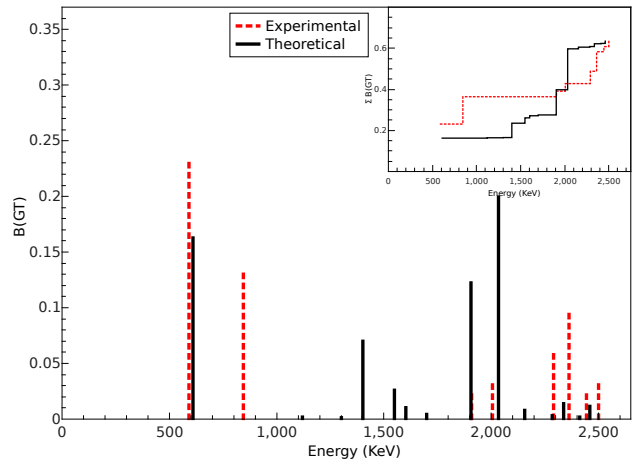}
 \caption{(Color online) Calculated $^{136}$Xe GT strengths (solid line) compared to the experimental ones (dotted line)\cite{exp_136xe}. The inset presents the calculated and the experimental GT running sum.}
 \label{gt_136xe}
 \end{figure}

The same value of the quenching factor observed for $pf$-shell nuclei \cite{cole-prc-2012}, $0.74$, was successfully used for the description of $^{136}$Xe \cite{hb_prl_2013}
in the larger $jj77$ model space.
As we have already mentioned in the beginning of Section \ref{detail}, the spin-orbit partners orbitals of $0g_{7/2}$ and $0h_{11/2}$ are missing in the $jj55$ model space, the Ikeda sum rule is 
not satisfied. This results in missing about half of the GT sum-rule, although the loss is at higher energies and cannot be seen in the insets of Fig. \ref{gt_130te} and Fig. \ref{gt_136xe}, which display the low-energy running GT strength sum. Both calculations use the same quenching factor, 0.74
 
Fig. \ref{gt_130te} presents our calculated GT strengths, plotted with solid black lines, for the transition of $^{130}$Te to $^{130}$I  compared to the 
experimental results shown with dotted red line \cite{exp_130te}. The running GT sum is displayed in the inset of the plot using the same line and colors convention as the GT strengths.
 Fig. \ref{gt_136xe} shows the calculated GT strengths of $^{136}$Xe to $^{136}$Cs compared to the experimental data \cite{exp_136xe}. 
 Also displayed in the plot inset is the running GT sum. The same color and lines convention as in Fig. \ref{gt_130te} has been used. Although there are discrepancies in the GT strength of individual states of these odd-odd nuclei, 
 the overall theoretical GT running sums are in reasonable agreement with the data.

\subsection{Analysis of the nuclear matrix elements for $\bf ^{136}$Xe} \label{nme-xe}

 Shell model methods for calculating double-beta decay NME have successfully predicted the correct $2\nu\beta\beta$ decay lifetime of $^{48}$Ca \cite{Cau90} prior to its experimental measurement.
% Nuclear matrix elements that were calculated with shell model methods have been successfully used to predict the correct $2\nu\beta\beta$ decay lifetime of $^{48}$Ca \cite{Cau90}.
In this case, the typical $pf$-shell nuclei \cite{cole-prc-2012} Gamow-Teller quenching factor, $0.74$, brings the calculated lifetimes within the experimental limits \cite{MH13}. In addition, the same quenching factor was  successfully 
used \cite{hb_prl_2013} to describe the $2\nu\beta\beta$ decay NME of $^{136}$Xe in the larger model space $jj77$ that contains all spin-orbit partner orbitals. Using this value of the quenching factor in the $jj55$ model space for the 
calculation of the $^{130}$Te and $^{136}$Xe $2\nu\beta\beta$ decay NME, we obtain $M^{2\nu\beta\beta}_{^{130}Te} = 0.0238 MeV^{-1}$ and $M^{2\nu\beta\beta}_{^{136}Xe} = 0.0256 MeV^{-1}$, respectively. The experimental results would indicate the need for slightly smaller
 values of the quenching factors: $qf_{^{130}Te}=0.59$, which leads to $M^{2\nu\beta\beta}_{^{130}Te} = 0.0175^{+0.0016}_{-0.0014} MeV^{-1}$ \cite{barabash-prc-2010} and 
$qf_{^{136}Xe}=0.71$ resulting in $M^{2\nu\beta\beta}_{^{136}Xe} = 0.0218 \pm 0.0003 MeV^{-1}$ \cite{exo-200}. This is clearly an artifact due to the missing spin-orbit partner orbitals in the $jj55$ model space. However, this effect seems to be 
smaller for the SVD Hamiltonian than when using other effective interactions (see e.g. Table 2 of \cite{plb2012} and $jj55$ restricted results in \cite{hb_prl_2013}). 

\begin{figure} 
 \includegraphics[width=0.95\linewidth]{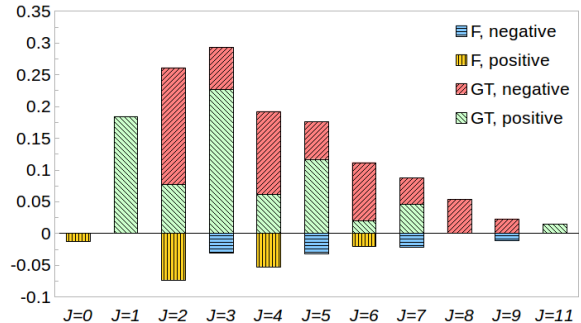}
 \caption{(Color online) $J_\kappa$-decomposition: contributions of the intermediate states $\left| \kappa \right > $ with certain spin and parity $J^\pi$ to the nonclosure
Gamow-Teller and Fermi matrix elements for the $0\nu\beta\beta$ decay of $^{136}$Xe (light neutrino exchange). CD-Bonn SRC parameterization was used.}
 \label{jk_decomp_light}
 \end{figure}

A beyond closure approach for the analysis of $^{130}$Te decomposition of the NME is currently out of reach due to the huge dimensions of the model space. Therefore, we test the SVD Hamiltonian for 
$^{136}$Xe in  
the smaller $jj55$ model space, for which we can perform several decompositions of the NME. 
Validation of the results for $^{136}$Xe using this interaction would make us confident in its reliable use for the calculation of the NME of $^{130}$Te. 
In addition, we use a recently proposed method \cite{sh_2014_prep} to calculate the optimal closure energy \cite{senkov-horoi-brown-prc2014} for $^{130}$Te by calculating the optimal closure energy for $^{136}$Xe, 
and we find $\left<E\right> = 3.5 \ MeV$. Similar to the analysis of $^{82}$Se in Ref.\cite{senkov-horoi-brown-prc2014}, we perform a beyond closure study of $^{136}$Xe NME.
Fig. \ref{jk_decomp_light} shows the Gamow-Teller and the Fermi (multiplied by the factor $( g_V/g_A) ^2$) beyond closure light neutrino exchange NME
calculated for a fixed spin and parity $J_\kappa^\pi$ of the intermediate states
 $\left| \kappa \right > $ \cite{sen-mh-prc88} ($J_\kappa$-decomposition). Having this $J_\kappa$-decomposition,
we find the total NME as a sum over all the spin contributions: $M_\alpha=\sum_{J_\kappa} M_\alpha (J_\kappa)$.
The relative sign of the Fermi matrix elements is opposite to that of the and Gamow-Teller matrix elements. Therefore, according to Eq. (3), the total size of each bar in Fig. \ref{jk_decomp_light} roughly corresponds to the total 
NME for a given $J_\kappa$ (the contributions of the Tensor NME are negligibly small, as seen in Table \ref{tab_nme}). 

\begin{figure} 
 \includegraphics[width=0.95\linewidth]{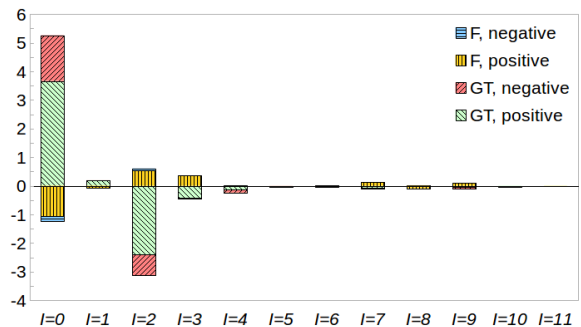}
 \caption{(Color online) $I$-pair decomposition: contributions to the running nonclosure
Gamow-Teller and Fermi matrix elements for the $0\nu\beta\beta$ decay of $^{136}$Xe (light neutrino exchange) from the configurations
when two initial neutrons and two final protons have a certain total spin $I$. CD-Bonn SRC parameterization was used.}
 \label{pair_decomp_light}
 \end{figure}

Another approach is the decomposition of the NME over the angular momentum $I$ of the proton (or neutron) pairs (see Eq. (B4) in Ref. \cite{sen-mh-prc88}) ($I$-pair decomposition). 
In this case, the NME can be written as $M_\alpha=\sum_{I} M_\alpha (I)$. Fig. \ref{pair_decomp_light}, presents this decomposition for the light neutrino exchange mechanism, 
where one can see the cancellation between $I=0$ and $I=2$, similar to the case of $^{82}$Se in Ref. \cite{senkov-horoi-brown-prc2014} and $^{48}$Ca in Ref. \cite{sen-mh-prc88}
 
\begin{figure} 
 \includegraphics[width=0.95\linewidth]{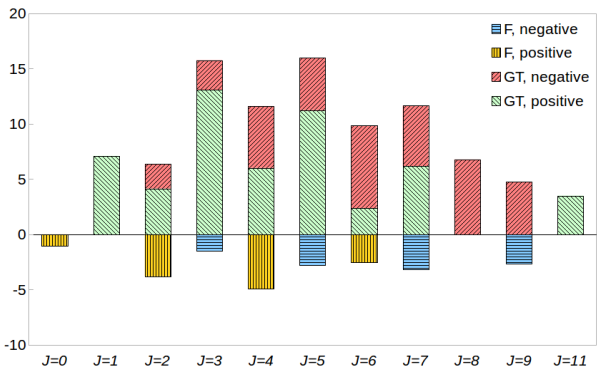}
 \caption{(Color online) Same as Fig. \ref{jk_decomp_light}, here for $^{136}$Xe heavy neutrino.}
  \label{jk_decomp_heavy}
 \end{figure}

\begin{figure} 
 \includegraphics[width=0.95\linewidth]{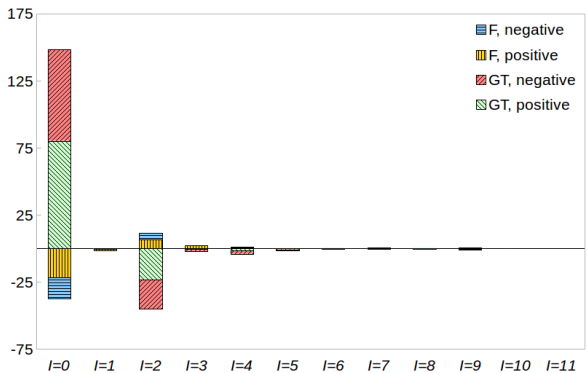}
 \caption{(Color online) Same as Fig. \ref{pair_decomp_light}, here for $^{136}$Xe heavy neutrino.}
 \label{pair_decomp_heavy}
 \end{figure}

 We also perform this analysis for the heavy neutrino exchange mechanism and we find a behavior similar to that of the light neutrino exchange mechanism. Fig. \ref{jk_decomp_heavy} 
 presents the the Gamow-Teller and the Fermi (multiplied by the factor $( g_V/g_A) ^2$) beyond closure heavy neutrino exchange matrix elements
calculated for a fixed spin and parity $J_\kappa^\pi$ of the intermediate states $\left| \kappa \right > $. Fig. \ref{pair_decomp_heavy} displays the $I$-pair decomposition for the heavy neutrino exchange mechanism.
 
 The numerical values for the $0\nu\beta\beta$ decay NME of $^{136}$Xe are presented in Table \ref{tab_nme}, where we show  the Gamow-Teller, the Fermi, and the Tensor contributions.
 The weak coupling strengths used in the calculations are $g_V=1$ and $g_A=1.254$. For both light and heavy neutrino exchange mechanisms, we use two recent SRC parameterizations (derived from the Argonne-V18 and CD-Bonn potentials \cite{sim-09}). 
 One can see that the effect of the choosing different SRC parameterizations on the value of the NME  is about 10\% in the case of the light neutrino 
  exchange mechanism, and about 30\% in the heavy neutrino exchange scenario (the transition operator is short-range).

\section{Nuclear matrix elements for $\bf ^{130}\text{Te}$} \label{nme-te}

%Having evaluated the SVD interaction against several experimental results concerning quadrupole electrical transitions, occupation probabilities, Gamow-Teller strengths, and having calculated
%the $^{136}$Xe nuclear matrix elements and disassembled the components according to their contributions in terms of parity and spin and the two-body matrix elements depending on the pairs 
%isospin, 
Having investigated the SVD Hamiltonian for $^{136}$Xe in the previous section, we calculate the NME for the $^{130}$Te.
Table \ref{tab_nme} presents the NME for $^{130}$Te alongside $^{136}$Xe, where two prescriptions for the short-range correlations, Argonne-V18 and 
CD-Bonn, were used. We show values for each contribution ($M^{0\nu}_{GT}$,$M^{0\nu}_{F} $ and $M^{0\nu}_{T}$), together with the total NME ($M^{0\nu}_{\nu}$) for both light and heavy neutrino exchange mechanisms.
%The value chosen for the $g_A$ strength is $1.254$, in order to easily compare our results with others results reported in the literature (using $g_A=1.269$ would decrease the NME by less than 1\%).
As one can see, the tensor contribution is negligible in the case of light neutrino exchange, while in the case of heavy neutrino exchange it is noticeable, but still very small. Also, in the 
case of $^{130}$Te and $^{136}$Xe, the tensor contribution leads to an increase in the value of the total NME. The following Section is dedicated to discussions on how the 
results presented here in Table \ref{tab_nme} for $^{130}$Te and $^{136}$Xe compare to other NME results reported in the literature.
\begin{table} 
\begin{ruledtabular}
\begin{tabular}{lccccc} 
			    & 		     	 &\multicolumn{2}{c}{$\nu$} & \multicolumn{2}{c}{$N$} \\
			    &    	     	 & AV18   & CD-Bonn & AV18    & CD-Bonn     \\ \hline 
\multirow{4}{*}{$^{130}$Te} &$M^{0\nu}_{GT}$ 	 & 1.54    & 1.66    & 70.76  	& 107.75     \\ 
			    &$M^{0\nu}_{F} $ 	 & -0.40   & -0.44   & -33.97	& -41.01       \\
			    &$M^{0\nu}_{T}$  	 & -0.01   & -0.01   & -2.24	& -2.24      \\
			    &$\bf M^{0\nu}_{\nu}$&\bf 1.80 &\bf 1.94 &\bf 94.60	&\bf 136.08      \\ \hline 
\multirow{4}{*}{$^{136}$Xe} &$M^{0\nu}_{GT}$	 & 1.39    & 1.50    & 63.53    & 96.68      \\
			    &$M^{0\nu}_{F} $	 & -0.37   & -0.40   & -30.64   & -36.95      \\
			    &$M^{0\nu}_{T}$	 & -0.01   & -0.01   & -2.42    & -2.42      \\
			    &$\bf M^{0\nu}_{N}$	 &\bf 1.63 &\bf 1.76 &\bf  85.43&\bf 122.59      \\ 
\end{tabular}
\end{ruledtabular}
 \caption{The NME for the light neutrino ($\nu$) and the heavy neutrino ($N$) exchange mechanism obtained with Argonne-V18 (AV18) and CD-Bonn
 SRC parameterizations, $\left<E\right>=3.5MeV$ and $g_A=1.254$.}
 \label{tab_nme}
\end{table}

\begin{figure}[htb] 
 \includegraphics[width=0.95\linewidth]{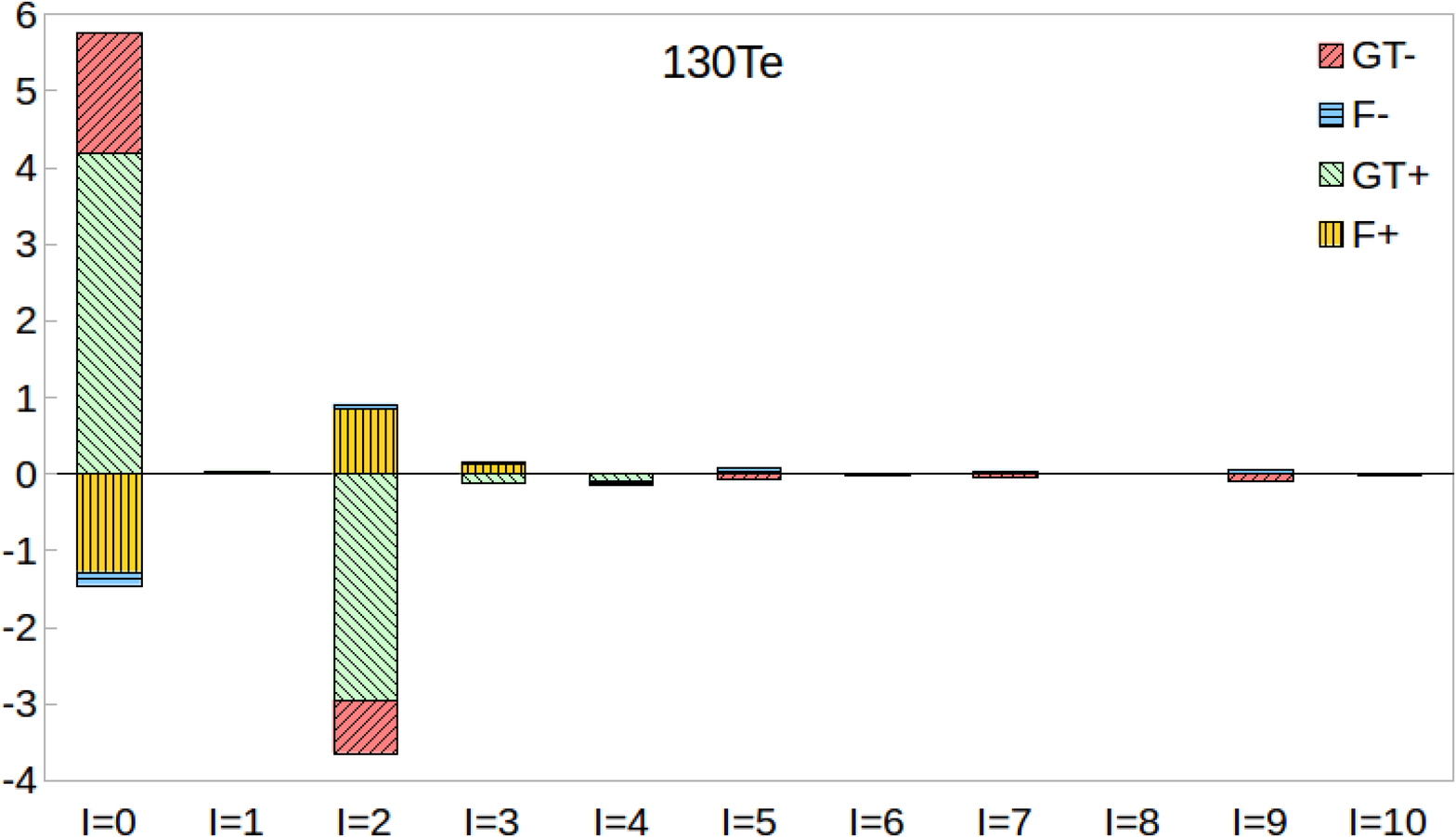}
 \caption{(Color online) Same as Fig. \ref{pair_decomp_light}, here for $^{130}$Te light neutrino closure NME decomposition.}
 \label{pair_decomp_te_light}
 \end{figure}
 
  \begin{figure}[htb] 
 \includegraphics[width=0.95\linewidth]{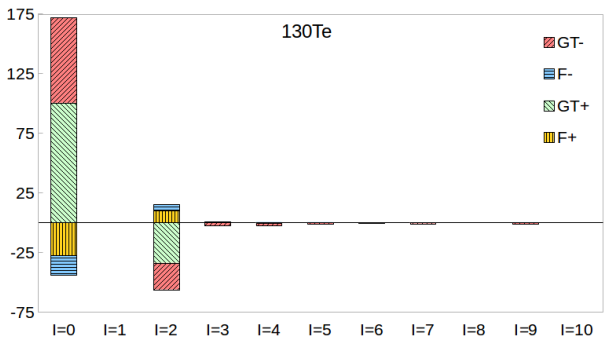}
 \caption{(Color online) Same as Fig. \ref{pair_decomp_light}, here for $^{130}$Te heavy neutrino closure NME decomposition.}
 \label{pair_decomp_te_heavy}
 \end{figure}

Similar to the $^{136}$Xe case, we perform an $I$-pair decomposition of the closure $^{130}$Te nuclear matrix elements. The optimal closure energy is 
$\left<E\right>=3.5MeV$. % and for the coupling strengths we have chosen the values $g_V=1$ and $g_A=1.254$. 
Figure \ref{pair_decomp_te_light} shows this decomposition of the NME for the light neutrino exchange mechanism. We observe that most of the contribution is from the pairs coupled to spin $I=0$ and $I=2$.
The Fermi components are multiplied by a factor $( g_V/g_A) ^2$, such that the size of each bar corresponds approximately
to the total NME value for a specific spin $I$ of the pairs.
Figure \ref{pair_decomp_te_heavy} presents the same decomposition of the NME for the heavy neutrino exchange mechanism. In this case the main contribution tot the total NME also comes from the pairs coupled to spin $I=0$ and $I=2$. This behavior seems to be universal, and it was recently used to propose an alternative method of calculating the NME \cite{bhs14} that could be further validated using information from pair transfer reactions.

\section{Comparison to other NME reported recently} \label{discussions}

\begin{figure}[htb] 
 \includegraphics[width=0.95\linewidth]{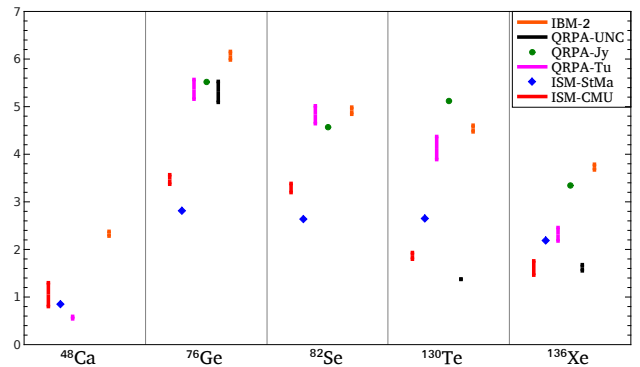}
 \caption{(Color online) Comparison of light neutrino exchange $0\nu\beta\beta$ NME obtained with different nuclear structure methods. Columns left to right correspond to down to up in the legend box.}
 \label{light}
 \end{figure}

\begin{figure}[htb] 
 \includegraphics[width=0.95\linewidth]{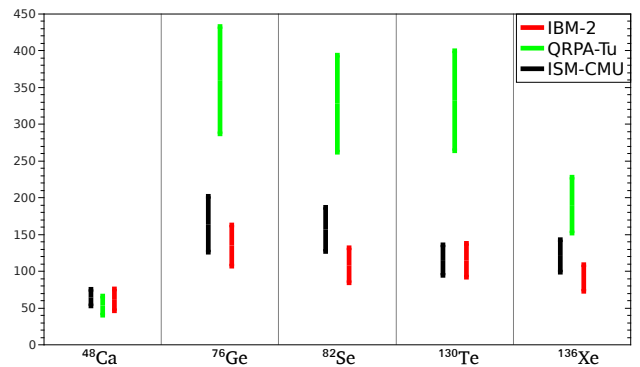}
 \caption{(Color online) Comparison of heavy neutrino exchange $0\nu\beta\beta$ NME obtained with different nuclear structure methods.Columns left to right correspond to down to up in the legend box.}
 \label{heavy}
 \end{figure}

Figs. \ref{light} and  \ref{heavy} show an overview of the most recent NME values calculated with ISM, QRPA and IBM-2 methods for five nuclei of immediate experimental interest and relevant for extracting neutrino properties. 
%In Fig. \ref{light} we present NME values for the light left-handed neutrino $0\nu\beta\beta$ decay and in Fig. \ref{heavy} the ones for the heavy right-handed neutrino $0\nu\beta\beta$ decay. 
. To keep our figures simple, we have only selected methods that includes correlations similar to the shell-model\cite{fp-cor}, and also provide results for the heavy neutrino case.
For the case of light neutrino, there are results from more methods, mentioned in the Introduction and presented e.g. in Fig. 6 of Ref. \cite{ring-bmf-2014}

Taken into account are NME calculated using softer short range correlation parameterizations extracted from Coupled Cluster Method (CCM) (based on Argonne-V18 and CD-Bonn potentials) \cite{sim-09},or Unitary Correlation
Operator Method (UCOM) \cite{UCOM}. In the case of the light neutrino exchange mechanism, the choice of SRC parameterization plays a smaller role in the final values of the NME, offering a variation usually up to 10\%, while in the heavy neutrino exchange scenario, 
its range is increased up to 30\%, due to the short-range nature of the transition operator. Regarding the influence of the $g_A$ strength value, in our calculations,
the differences in the NME values are less than half percent when changing from the usual value of 1.254 to the latest experimentally determined value of 1.269 \cite{iba-prl12}.
%For a consistent analysis, we report and compare the results obtained with $g_A=1.254$.

We also notice the new light neutrino QRPA results for $^{130}$Te and $^{136}$Xe reported in \cite{qrpa-en-2013}, which are very close to our shell model calculations. 
As a general trait, the ISM light neutrino results are different, usually by a factor of two, from other methods. In the case of the heavy neutrino, our shell model results are much closer to the IBM-2 ones, but still different
from QRPA calculations by a factor of two. Due to the short-range nature of the heavy neutrino operator, we find an increased dependency of the results on the SRC parameterization employed.
 
 \begin{table*} [htb]
  \caption{The calculated light neutrino $0\nu\beta\beta$ decay NME obtained with different nuclear structure methods. Two NME values separated by the slash sign denote the results obtained
  with Argonne-V18 and with CD-Bonn SRC, respectively. A single NME value means that it was calculated using UCOM SRC.}
\begin{ruledtabular}
 \begin{tabular}{lccrcccc} 
	&$^{48}$Ca	&$^{76}$Ge	&$^{82}$Se	&$^{130}$Te	&$^{136}$Xe	\\ \hline 
IBM-2   &2.28/2.38	&5.98/6.16	&4.84/4.99	&4.47/4.61	&3.67/3.79	\\									
QRPA-UNC&		&5.09/5.53	&		&1.37/1.38	&1.55/1.68	\\ 
QRPA-Jy &		&5.52		&4.57		&5.12		&3.35	\\ 
QRPA-Tu &0.54/0.59	&5.16/5.56	&4.64/5.02	&3.89/4.37	&2.18/2.46	\\ 
ISM-StMa&0.85		&2.81		&2.64		&2.65		&2.19	\\ 
ISM-CMU &0.80/0.88	&3.37/3.57	&3.19/3.39	&1.79/1.93	&1.63/1.76	\\ 
 \end{tabular}
 \end{ruledtabular} 
 \label{tab_nme_comp_light}
 \end{table*}
 
In Fig. \ref{light} we show a comparison of the different light neutrino NME recently reported by different groups using the most recent update of their calculation. 
ISM-CMU are the shell model results of our group obtained with Argonne-V18 and CD-Bonn SRC. For $^{48}$Ca, the results are taken from \cite{sen-mh-prc88} and 
\cite{48ca-high-prec}. The $^{76}$Ge NME are from \cite{sh_2014_prep}, while the $^{82}$Se results are from \cite{senkov-horoi-brown-prc2014}.
The $^{130}$Te values are from this work. $^{136}$Xe NME results are from \cite{hb_prl_2013} and from Table \ref{tab_nme}.
ISM-StMa denote the interacting shell model results of the Strasbourg-Madrid group published in \cite{npa818}, which were obtained using UCOM SRC. QRPA-Tu are the QRPA results of the Tuebingen group and the NME 
are selected from their very recent paper \cite{qrpa-tu-2014}, for Argonne-V18 and CD-Bonn SRC parameterizations.
QRPA-Jy represent the QRPA calculations of the Jyvaskyla group, and their results are taken from \cite{qrpa-jy-2010}, where SRC is taken into account using UCOM. 
QRPA-UNC show the QRPA results of the University of North Carolina group \cite{qrpa-en-2013}, where SRC were omitted. 
The IBM-2 are NME from the Yale group \cite{iba-prc2013}, for Argonne-V18 and CD-Bonn SRC parameterizations.

 Fig. \ref{heavy} presents the heavy neutrino NME obtained with three different methods. The results use the same conventions and parameters as those presented in Fig. \ref{light}. 
ISM-CMU results are from Ref. \cite{sen-mh-prc88} for $^{48}$Ca, Ref. \cite{sh_2014_prep} for $^{76}$Ge, Ref. \cite{senkov-horoi-brown-prc2014} for $^{82}$Se, 
present work for $^{130}$Te, and present work and \cite{hb_prl_2013} for $^{136}$Xe. QRPA-Tu are results from \cite{qrpa-tu-2014},
and IBM-2 are results from \cite{iba-prc2013}.

 Table \ref{tab_nme_comp_light} shows the NME displayed in Fig. \ref{light} for the the light neutrino exchange. The references and notations correspond to those in the figure.
 Displayed in Fig. \ref{light}, but not presented in Table \ref{tab_nme_comp_light}, are the ISM-CMU values 1.30 for $^{48}$Ca \cite{48ca-high-prec} obtained with an effective $0\nu\beta\beta$ operator, and 1.46 
 for $^{136}$Xe \cite{hb_prl_2013} calculated in the larger $jj77$ model space.

   \begin{table} [htb]
 \caption{The calculated heavy neutrino $0\nu\beta\beta$ decay NME obtained with different nuclear structure methods.The two NME values separated by the slash sign denote the results obtained
  with Argonne-V18 and with CD-Bonn SRC, respectively.}
  \begin{ruledtabular}
 \begin{tabular}{@{}l@{}ccccccc@{}} 
	       &$^{48}$Ca		&$^{76}$Ge	&$^{82}$Se	&$^{130}$Te	&$^{136}$Xe	\\ \hline 
\small IBM-2   &46.3/76.0		&107/163	&84.4/132	&92.0/138	&72.8/109	\\									
\small QRPA-Tu &40.3/66.3		&287/433	&262/394	&264/400	&152/228	\\ 
\small ISM-CMU \  &52.9/75.5		&126/202	&127/187	&94.5/136	&98.8/143	\\ 
 \end{tabular}
  \end{ruledtabular} 
 \label{tab_nme_comp_heavy}
 \end{table}
 
In Table \ref{tab_nme_comp_heavy} we list the NME displayed used in Fig. \ref{heavy} for the the heavy neutrino exchange. The choice of parameterizations and ingredients is identical to the 
one in Table \ref{tab_nme_comp_light}. There are less reported results for the heavy neutrino exchange mechanism than for the light neutrino exchange. for reasons explained above we only present the results obtained with IBM-2, 
QRPA and ISM methods. Also, in this scenario, we encounter the largest uncertainties and variations in the NME values, mainly due to the stronger dependency on the SRC parameterization used in the calculations.
%Since in this case, the IBM-2 results are very close to the SM ones, it would be interesting to see how the NME values from other methods would compare to these or to the QRPA ones, should they
%be evaluated and reported.

\section{Conclusions} \label{conclusions}

In this paper, we calculate the $0\nu\beta\beta$ decay NME for $^{130}$Te in the $jj55$ model space. We use an effective Hamiltonian that was fine-tuned with experimental data, reported in Ref. \cite{svd_int}.
Two decay mechanisms are considered: the light left-handed neutrino exchange and the heavy right-handed neutrino exchange. We report new $^{130}$Te NME in the range $1.79-1.93$ for light neutrino 
and $94.5-136$ for heavy neutrino.

Additional to the new NME results, we provide a detailed analysis and validation of the effective Hamiltonian by performing calculations of several spectroscopic quantities 
(energy spectra, $B(E2)\uparrow$ transitions, occupation probabilities and $GT$ strengths) and comparing them to the available experimental data. 
We find good agreement with the data for the nuclei of interest.

We perform calculations of the $2\nu\beta\beta$ decay NME of $^{130}$Te and $^{136}$Xe, where we find newer quenching factors that bring the calculations within the experimental result limits. 
These values are higher than the ones previously proposed for the $jj55$ model space in Ref. \cite{plb2012}. When comparing to the more complex calculations performed within a 
larger model space (the $jj77$) from Ref. \cite{hb_prl_2013}, we obtain slightly larger $2\nu\beta\beta$ and $0\nu\beta\beta$ decay NME. 
This is an effect attributed to the missing spin orbit partners in the $jj55$ model space.

We present the analysis of the $^{136}$Xe nonclosure NME, where we investigate the $J_\kappa$-decomposition and the $I$-pair decomposition. 
Also provided, is an analysis of the $I$-pair decomposition for the $^{130}$Te closure NME.
In both nuclei, the main contribution to the NME value is identified as arising from the cancellation between $I=0$ and $I=2$ pairs, similar to the case of $^{48}$Ca \cite{sen-mh-prc88} and $^{82}$Se \cite{senkov-horoi-brown-prc2014}.

We also present an overview of the NME for both light neutrino and heavy neutrino exchange reported in the literature, calculated using different methods. 
We noticed the new QRPA-UNC light neutrino results for $^{130}$Te and $^{136}$Xe that are close to our ISM results. In the case of heavy neutrino, we get similar results to IBM-2.
 
\begin{acknowledgments}
The authors thank B.A. Brown and R. Senkov for useful discussions and advise.
Support from  the NUCLEI SciDAC Collaboration under
U.S. Department of Energy Grants No. DE-SC0008529 and DE-SC0008641 is acknowledged.
MH also acknowledges U.S. NSF Grant No. PHY-1404442\end{acknowledgments}

% Create the reference section using BibTeX:

\end{document}